\documentclass[journal]{IEEEtran}
\usepackage[english]{babel}
\usepackage[utf8]{inputenc}
\usepackage[pdftex]{graphicx}
\usepackage{amsmath,mathrsfs}
\usepackage{algorithmic}
\usepackage{array}
\usepackage{mdwmath}
\usepackage{mdwtab}
\usepackage{eqparbox}
\usepackage[tight,footnotesize]{subfigure}
\usepackage[pdftex,usenames,dvipsnames]{color}
\usepackage{siunitx}
\usepackage{listings}
\usepackage[hidelinks]{hyperref}
\usepackage{parskip}
\usepackage{color}
\usepackage{comment}
\usepackage{braket}

\definecolor{mygreen}{rgb}{0,0.6,0}
\definecolor{mygray}{rgb}{0.5,0.5,0.5}
\definecolor{mymauve}{rgb}{0.58,0,0.82}
\begin{document}

\title{
	\Huge \textbf{Obtaining the Drude Equation for Electrons in Metals Using a Fractional Variational Principle} \vspace{0.5cm}
}
\date{\today}
\author{
  \centering
  \begin{tabular}{c c}
  Luis Fernando Mora Mora & B24449 
  \end{tabular}
}
\author{Luis Fernando Mora Mora\\
        luis.mora\_m@ucr.ac.cr\\
        Universidad de Costa Rica}%

\maketitle

\begin{abstract}
A fractional variational principle was derived in order to be used with lagrangians containing fractional derivatives of order 1/2. By forcing the action associated to this type of lagrangian to be stationary, a modified fractional Euler-Lagrange equation was obtained. This was shown to reproduce the equations of motion of two basic 1-dimensional energy-dissipative systems: a spring-mass system damped by friction, and a RLC circuit connected in series. Finally, by using the fractional Euler-Lagrange equation, the Drude relationship for electrons in metals was recovered when a fractional kinetic energy was taken into consideration in the electron's associated energies.
\end{abstract}



\section{Introduction}

\subsection{Fractional Calculus and Lagrangian Mechanics}

Fractional calculus refers to integration and derivation of functions in non-integer orders, and is as old as its classical counterpart, integer-order calculus \cite{herrmann2011fractional}. Just as Euler's Gamma function allows the generalization of a factorial function to any real or complex number, it also allows the generalization of any integral and derivative to arbitrary order through the means of Cauchy's integral \cite{li2015numerical,herrmann2011fractional,guia2016fractional}.\\

During its origins, which date to Leibniz, fractional calculus was developed in a slow way, being studied mostly only by mathematicians as an abstract branch of mathematics with little or no application \cite{li2015numerical}. Nonetheless, starting from the 1970's, its development has accelerated and numerous applications of fractional calculus to physics have been found, and attracted the attention of many researchers and engineers. These applications include: materials and proccesses with long-term memory, quark confinement, anomalous diffusion in plasmas, long range interactions, and fractals to name a few \cite{li2015numerical,drummond1962anomalous,herrmann2011fractional,tatom1995relationship}.\\

Similarly, based on integer-order calculus, lagrangian and hamiltonian analysis are the foundations of classical physics. Lagrangian mechanics started with the work of Joseph-Louis Lagrange, who derived a mathematical formalism through the means of calculus of variations, which is capable of reproducing Newton's laws of motion for a system based solely on its associated energies \cite{taylor2005classical, goldstein2002classical,greiner2009classical}. By forcing the action of a system to be stationary, a set of equations called the \textit{Euler-Lagrange} equations can be obtained, which describe the temporal evolution of a system described by a set of generalized coordinates $q$:

\begin{equation}
    \frac{d}{dt}\Big(\frac{\partial \mathcal{L}}{\partial \dot{q_i}}\Big) = \frac{\partial \mathcal{L}}{\partial q_i}
\end{equation}

This method provides the equations of movement for a conservative system without using force vectors, as in Newtonian analysis. This formalism is widely known as the \textit{principle of least action } and has been widely studied in all kinds of systems in classical mechanics \cite{taylor2005classical,greiner2009classical,landau1982mechanics,goldstein2002classical}. 

Nonetheless, when classical lagrangian analysis is applied to disipative systems, the least action principle fails. As shown by Bauer, the equations of movement of an energy-disipative system can't be obtained through the use of a variational principle \cite{bauer1931dissipative, riewe1996nonconservative}. In order to analyze disipative systems through a variational principle, it is necessary to turn to the tools fractional calculus has to offer.

Fractional operators act as a ``legal loop-hole'' in Bauer's theorem, which only takes into account inter-order integration and derivation. This exception has been put into practice by considering lagrangians with fractional energies \cite{dreisigmeyer2003nonconservative, SGil,HamiltronFractional}, which yield a modified set of Euler-Lagrange equations.

In this work, a fractional approach towards mechanical and electrical systems is developed. Now we proceed to introduce and define some crucial properties of fractional operators. 

\subsection{Properties of Fractional Derivatives}

There are at least six ways to define a fractional derivative, which are not equivalent to each other \cite{li2015numerical, herrmann2011fractional,loverro2004fractional}. Even though there exists a great diversity of definitions, the Riemann-Louiville definition will be cited here, even though they all share similar group properties. Consider a time interval $a<t$, where $a$ represents an arbitrary parameter. Next, consider a system with a generalized coordinate $q(t)$ such that we wish to compute its $\alpha$-order derivative with respect to $t$, with $\alpha$ a positive and real number. For $m$ the first integer greater than $\alpha$, the so-called \textit{right hand} $\alpha$-order derivative for \textit{q} is given by \cite{herrmann2011fractional,SGil,li2015numerical}:  

\begin{equation}
    \label{eq:causalderivative}
     {}^{}_{a}D^{\alpha}_{t}[q(t)] = \frac{d^m}{dt^m}\Big[ \frac{1}{\Gamma(m-\alpha)}\int_a^t q(\tau)(t-\tau)^{m-\alpha-1}d\tau\Big]
\end{equation}

Where $\Gamma(\cdot)$ is Euler's gamma function. To simplify our notation, any inter-order derivative will be written with the usual dot notation, but any fractionary derivative will be written as:

  \begin{equation}\label{ar13}  
    _aD^{\alpha}_{t}[q(t)] = q^{(\alpha)}
  \end{equation}

Note that this type of derivative, being computed through an integral, implies that the function is non-local in time (it's not pointwise, one requires the entire history of the function until time $t$ to be able to compute it). For this reason, it is said that it has a \textit{\textbf{memory}} property. This property tells us how a system can be affected by its past, at that memory vanishes, such that the function is more affected by its recent past than its distant past \cite{SGil,dreisigmeyer2003nonconservative}. Also, fractionary operators are linear, and satisfy the additive index law \cite{rudolf2000applications,li2015numerical,herrmann2011fractional}: 

\begin{equation}
    \Big[(q+p)^{\alpha}\Big]^{\beta} = q^{(\alpha+\beta)}+p^{(\alpha+\beta)}
\end{equation}

Leibniz's rule for the derivative of a product does not apply to the fractional case. If one wishes to compute the fractionary derivative of the product of two functions, one must evaluate the following expression for $j<\alpha$:

\begin{equation}
    (uv)^{(\alpha)} = \sum_{j=0}^{\infty} \binom{\alpha}{j}u^{(\alpha-j)}v^{(j)} 
\end{equation}

\begin{figure}[h!]
\centering
\includegraphics[scale=0.5]{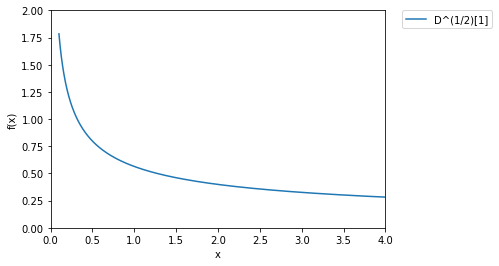}
\caption{1/2-order derivative of A = 1}
\label{adf}
\end{figure}

For example, under this definition, the fractionary derivative of a constant is not cero. For a constant $A$, its 1/2-order derivative is shown in equation 5 and shown in figure \ref{adf}: 

\begin{equation}
    A^{(1/2)} = \frac{A}{\Gamma(1/2)}(t-a)^{-1/2}
\end{equation}

A useful tool to understanding fractional derivatives lies in Fourier analysis. Its possible to prove that the Fourier transform of a Riemann-Louville fractionary derivative fulfills the following property analogous to its inter-order counterpart:

\begin{equation}
    \mathscr{F}[_aD_t^\alpha[q(t)]] = \omega^\alpha e^{i\alpha\pi/2}\tilde{q}(\omega)
\end{equation}

With $\tilde{q}(\omega)$ the Fourier transform of \textit{q(t)}. As equation 7 shows, taking the $\alpha$-order derivative of a function causes a rotation of $\alpha\pi/2$ radians counterclockwise and an amplification by a factor $\omega^\alpha$ for every component of frequency $\omega$ in the complex plane.


\section{Fractional Variational Principle}

Using the properties described for fractionary operators, we wish to obtain the equations of movement of an energy-dissipative system through the means of a variational principle, analogous to those of equation 1 which are the original Euler-Lagrange's equations. Consider the lagrangian of an arbitrary system, that depends on its generalized coordinate $q$, its first order derivative, or velocity, $\dot{q}$, but that also depends on the 1/2-order derivative of q:

\begin{equation}
    \mathcal{L} = \mathcal{L}(\dot{q},q^{(1/2)},q)
\end{equation}

So the problem we face is to make the action associated to this lagrangian to be stationary:

\begin{equation}
  \delta S =  \delta \int_a^b \mathcal{L}(\dot{q},q^{(1/2)},q) dt = 0
\end{equation}

Here a and b are arbitrary time parameters. To solve this variation  \cite{taylor2005classical}, an erroneous trajectory $Q(t)$ is taken with a small constant parameter $\gamma$ such that:

\begin{equation}
    Q(t) = q(t) + \eta(t)\gamma
\end{equation}

Also, the trajectory $\eta$ is chosen such that it satisfies:

\begin{equation}
    \eta(a) = \eta(b) = 0
\end{equation}

This redefines the action to be of the form:

\begin{equation}
   S(\gamma) = \int_a^b \mathcal{L}(\dot{q}+\dot{\eta}\gamma,q^{(1/2)}+\eta^{(1/2)}\gamma,q+\eta\gamma) dt
\end{equation}

Using a first-order Taylor expansion for $\mathcal{L}$, the variation with respect to the paramter $\gamma$ can be expressed as:

\begin{equation}
    \frac{dS}{d\gamma} = \int_a^b\Big( \frac{\partial \mathcal{L}}{\partial \dot{q}}\dot{\eta} + \frac{\partial \mathcal{L}}{\partial q^{(1/2)}}\eta^{(1/2)} + \frac{\partial \mathcal{L}}{\partial q}\eta\Big) dt = 0
\end{equation}

This integral is very similar to that encountered in classical mechanics \cite{taylor2005classical,goldstein2002classical}, and can be solved through the means of integration by parts. Consider first the $\dot{\eta}$ dependent term and integrate by parts:

\begin{equation}
 \int_a^b  \frac{\partial \mathcal{L}}{\partial \dot{q}}\dot{\eta} dt = \Big[\frac{\partial \mathcal{L}}{\partial \dot{q}}\eta\Big]\Big|_a^b - \int_a^b \frac{d}{dt}\Big(\frac{\partial \mathcal{L}}{\partial \dot{q}}\Big) dt  
\end{equation}

By the condition in equation 11, the boundary term is equal to cero, so the integral reduces to:

\begin{equation}
 \int_a^b  \frac{\partial \mathcal{L}}{\partial \dot{q}}\dot{\eta} dt = - \int_a^b \frac{d}{dt}\Big(\frac{\partial \mathcal{L}}{\partial \dot{q}}\Big)\eta dt  
\end{equation}

As mentioned before, this is a widely known result in classical mechanics that appears when deriving the original Euler-Lagrange equations \cite{taylor2005classical, goldstein2002classical, greiner2009classical,landau1982mechanics}. Now we wish to use the same integration by parts for the fractionary term. In order to use the modified Leibniz rule, a little more work is needed. First, let u and v be any two well-behaved functions, whose product shall have a fractionary derivative according to equation 5 given by:

\begin{equation}
    (uv)^{(1/2)} = u^{(1/2)}v
\end{equation}

\begin{equation}
    (vu)^{(1/2)} = v^{(1/2)}u
\end{equation}

Now sum the last two equations to obtain:

\begin{equation}
    (uv + vu)^{(1/2)} = u^{(1/2)}v + uv^{(1/2)}
\end{equation}

This last expression being equivalent to:

\begin{equation}
    uv^{(1/2)} = (uv + vu)^{(1/2)} - u^{(1/2)}v
\end{equation}

Applying equation 19 to the fractionary part of the integrand we get:

\begin{equation}
\frac{\partial \mathcal{L}}{\partial q^{(1/2)}}\eta^{(1/2)} = \Big[2\frac{\partial \mathcal{L}}{\partial q^{(1/2)}}\eta\Big]^{(1/2)} - \frac{d^{(1/2)}}{dt^{(1/2)}}\Big(\frac{\partial \mathcal{L}}{\partial q^{(1/2)}} \Big)\eta dt 
\end{equation}

For the first term on the right side of equation 20, the integration can be taken as a composition of two 1/2-order integrals:

\begin{equation}
\int_a^b \Big[2\frac{\partial \mathcal{L}}{\partial q^{(1/2)}}\eta\Big]^{(1/2)} dt = I^{(1/2)
}I^{(1/2)}\Big[2\frac{\partial \mathcal{L}}{\partial q^{(1/2)}}\eta\Big]^{(1/2)}
\end{equation}

But the first 1/2-order integral evaluates to cero by the condition of equation 11:

\begin{equation}
    \Big[2\frac{\partial \mathcal{L}}{\partial q^{(1/2)}}\eta\Big]\Big|_a^b = 0
\end{equation}

So finally the integration by parts for the fractionary integrand yields:

\begin{equation}
\int_a^b \frac{\partial \mathcal{L}}{\partial q^{(1/2)}}\eta^{(1/2)}dt = - \int_a^b \frac{d^{(1/2)}}{dt^{(1/2)}}\Big(\frac{\partial \mathcal{L}}{\partial q^{(1/2)}} \Big)\eta dt 
\end{equation}

Lastly, substitute equations 15 and 23 in equation 13, and by factorizing $\eta$ the condition for stationary action is obtained:

\begin{equation}
\int_a^b \Big[ - \frac{d}{dt}\Big(\frac{\partial \mathcal{L}}{\partial \dot{q}}\Big) - \frac{d^{(1/2)}}{dt^{(1/2)}}\Big(\frac{\partial \mathcal{L}}{\partial q^{(1/2)}} \Big) + \frac{\partial \mathcal{L}}{\partial q} \Big]\eta dt = 0 
\end{equation}

Because the action must be stationary for any trajectory $\eta(t)$, the term between brackets must be equal to cero, in order for the variation to be cero. This gives us the modified or \textit{fractionary Euler-Lagrange equations} for this problem:

\begin{equation}
    \frac{d}{dt}\Big(\frac{\partial \mathcal{L}}{\partial \dot{q}}\Big) + \frac{d^{(1/2)}}{dt^{(1/2)}}\Big(\frac{\partial \mathcal{L}}{\partial q^{(1/2)}} \Big) = \frac{\partial \mathcal{L}}{\partial q}
\end{equation}

Note that this modified version obtained is mostly equal to the original shown in equation 1, but a fractionary term appears, as should be expected from the addition of a fractionary derivative in the lagrangian.

\section{Fractional Variational Principle Applied}

Next it will be shown that the result of equation 25 is capable of reproducing the equations of motion for two basic, 1-dimensional dissipative system, one electrical and one mechanical.

\begin{figure}[htbp!]
    \centering
    \includegraphics{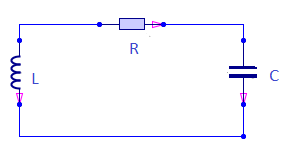}
    \caption{RLC circuit connected in series}
    \label{fig:RLC}
\end{figure}

First consider a RLC circuit connected in series, such as the one shown in figure \ref{fig:RLC}. In this circuit, R is the resistor's resistance, L is the inductor's inductance and C capacitor's capacitance.

For this circuit, let the charge $q$ stored in the capacitor be the generalized coordinate of the system. The inductor is going to store a magnetic energy $W_L = \frac{1}{2}L\dot{q}^2$, and the capacitor is going to store an electrostatic potential energy $W_e = q^2/2C$. Now for the resistor, a fractionary kinetic energy is introduced:

\begin{equation}
    W_R = \frac{1}{2}Rq^{(1/2)^2}
\end{equation}

Note that the 1/2-order derivative of the charge q is squared. This can be thought of as a fractionary ``resistance energy''. With these three energies in mind, the following lagrangian can be formulated:

\begin{equation}
    \mathcal{L} = \frac{1}{2}L\dot{q}^2 + \frac{1}{2}Rq^{(1/2)^2} - \frac{q^2}{2C}
\end{equation}

And by applying the fractionary Euler-Lagrange equation to this lagrangian, the differential equation for this circuit is obtained \cite{dorf2000circuitos} without using any Kirchoff laws:

\begin{equation}
    L\ddot{q} + R\dot{q} + \frac{1}{C}q = 0 
\end{equation}

\begin{figure}[htbp!]
    \centering
    \includegraphics{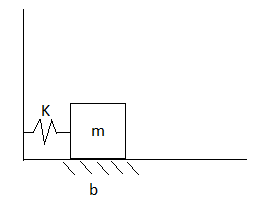}
    \caption{Spring-mass system with viscous damping coefficient b}
    \label{fig:masa}
\end{figure}

Analogously, the equations of movement for a spring-mass system with friction can be obtained with this method. Such system is shown in figure \ref{fig:masa}. 

Let the displacement from the spring's equilibrium position be the generalized coordinate of the system $q$. Similarly to the RLC circuit, a fractionary kinetic energy is introduced for the spring-mass system's lagrangian:

\begin{equation}
        \mathcal{L} = \frac{1}{2}m\dot{q}^2 + \frac{1}{2}bq^{(1/2)^2} - \frac{1}{2}Kq^2
\end{equation}

Where $m$ is the block's mass, $K$ the spring's constant and $b$ is an arbitrary viscous damping coefficient, also called friction coefficient. The coefficient $b$ varies with each system, and is chosen to fit the system's impulse-response curves. In comparison with its electrical counterpart, the viscous coefficient $b$ plays the same role as the resistance R in the RLC circuit. When the fractionary Euler-Lagrange equation is applied to this lagrangian, one obtains the equations of movement for the system:

\begin{equation}
    m\ddot{q} + b\dot{q} + Kq = 0 
\end{equation}

\section{Drude Model for Metals}

At the beginning of the XX century, Paul Drude described for the first time the movement of electrons inside metals using Boltzmann's kinetic theory of gases. His theory provided basic but successful principles in understanding metallic conduction, though sadly, he did not live to see his theory become influencial \cite{simon2013oxford}. In Drude's theory, the following assumptions are made about electron movement: 

\begin{itemize}
    \item Electrons experience a mean time $\tau$ between collisions, such that the probability of a collision within a time interval $dt$ is $dt/\tau$
    \item When scattering/colliding, the electron moves to a state of \textbf{average} momentum \textbf{p} = 0. 
    \item Electrons with charge -e respond to any electric or magnetic fields applied in the moments in between collision events. 
\end{itemize}

By these assumptions, if at time $t$ the average momentum of the electron is \textbf{p}(t), we wish to know what will be its value after a time $t+dt$. Since there is a probability $dt/\tau$ of a collision, complementarily there is a probability $1-dt/\tau$ that it will not scatter and will accelerate in response to any force applied. So by using the impulse theorem the following relationship is obtained:

\begin{equation}
    \textbf{p}(t+dt) = (1-dt/\tau)(\textbf{p}(t)+Fdt)+\textbf{0}dt/\tau
\end{equation}

By keeping only the linear terms in $dt$ the latter is equivalent to:

\begin{equation}
    \textbf{p}(t+dt) = \textbf{p}(t)+Fdt - \frac{\textbf{p}dt}{\tau}
\end{equation}

And by a little rearrengement this gives:

\begin{equation}
    \frac{\textbf{p}(t+dt)-\textbf{p}(t)}{dt} = F - \frac{\textbf{p(t)}}{\tau}
\end{equation}

By recognizing the definition of the first derivative for \textbf{p} on the left side of the equation, this leads to the famous Drude equation for metals:

\begin{equation}
    \frac{d\textbf{p}}{dt} = F - \frac{\textbf{p}}{\tau}
\end{equation}

Usually the term $-\textbf{p}/\tau$ is interpreted as a ``drag force'' on the electron, like it was inmersed in a viscous fluid. This type of dissipative force is now our focus of study and we intend to re-derive it through the formalism of fractional lagrangians.

\subsection{Fractional Variational Principle Applied to the Electron}

The Drude model for electrons involves a drag force which is similar to the friction experienced by the block mass of figure 3. We wish to find a fractionary lagrangian that will reproduce this effect for the electron, without invoking the three statistical assumptions used in Drude's theory. 

Let the same electron be considered, with average momentum \textbf{p} and average position \textbf{q}, and let it be subject to a generalized potential V(\textbf{q}). The fractionary lagrangian to be considered is proposed to have the form: 

\begin{equation}
\mathcal{L} = \frac{1}{2}m\dot{\textbf{q}}^2 + \frac{1}{2}\frac{m}{\tau}\textbf{q}^{(1/2)^2} - V    
\end{equation}

Note that the first term is just the kinetic energy of the electron, the last term corresponds to its potential energy, but the middle term here introduced is a fractional kinetic energy analogous to the one used in the spring-mass system. Here the arbitrary viscous coefficient $b$ is no longer arbitrary but proportional to the electron's mass:
\begin{equation}
 b \xrightarrow{} m/\tau   
\end{equation}
Since this fractionary kinetic energy varies with the scattering time $\tau$, it will vary according to the metal taken into consideration. When the fractional Euler-Lagrange equations are applied to the lagrangian of equation 35 we get:

\begin{equation}
    m\ddot{\textbf{q}} + \frac{m}{\tau}\dot{\textbf{q}} = -\nabla V
\end{equation}

Now, substituting for $\textbf{p} = m\dot{\textbf{q}}$ the last expression becomes:

\begin{equation}
    \dot{\textbf{p}} + \frac{1}{\tau}\textbf{p} = -\nabla V
\end{equation}

Finally, taking $F = -\nabla V$ the Drude equation for metals is recovered:

\begin{equation}
    \frac{d\textbf{p}}{dt} = F - \frac{\textbf{p}}{\tau}
\end{equation}

\section{Conclusions}

The fractional variational principle developed in this work was shown to give the equations of motion for two very basic dissipative systems: a spring-mass system damped by friction and a RLC circuit in series. This opens the possibility of using the modified Euler-Lagrange equations in the study of other dissipative systems of greater complexity as a next step in the development of fractional lagrangian mechanics.  

Secondly, it was shown that it is possible to obtain the Drude equation for electrons in metals through the means of the same fractional variational principle, without using the statistical approach used by Drude. A lagrangian containing a fractional kinetic energy that varies with $m/\tau$ was used to obtain the Drude equation, and consequently, this energy will depend on the material considered, as $\tau$ varies with each metal. 

By the results obtained, the fractional variational principle has proven to be an alternative method towards understanding friction, electrical resistance and electron conduction. A hamiltonian operator very similar to the electron's lagrangian of equation 35 was obtained in a past article \cite{mora2019link}, where $\tau = 1$. The similar result obtained for the electron provides a new relationship to be considered in the fractionary momentum paradigm for quantum particles.
 
As a final remark, from the RLC's resistive energy and the Drude electron's fractional kinetic energy, the following question arises: ¿is superconduction phenomena related to  minimizing or cancelling this fractional kinetic energy? Note that the problem of minimizing it is more complex than just taking an infinite relaxation time $\tau\xrightarrow{}\infty$ as one may naively suggest as a first alternative. To adress this question, further work is needed to couple the fractionary formalism of 1/2-order momentum operators to superconduction theory (BCS theory) which is essentially quantum mechanical.

\bibliographystyle{IEEEtran}
\bibliography{bibliography}

\end{document}